\documentclass[fleqn,12pt]{wlscirep}
\usepackage[utf8]{inputenc}
\usepackage[T1]{fontenc}
\usepackage{graphicx}
\usepackage{caption}
\usepackage[tableposition = top]{caption}
\usepackage[figureposition = top]{caption}
\usepackage[position=top]{subfig}
\title{Compensation of Absorption Effects in Seismic Data}

\author[1,*]{Gebregergs, Hagos Gebrehiwet}
\author[2]{Leiv, Jacob Gelius}
\affil[1]{Mekelle University,  Department of Physics, Mekelle, 231, Ethiopia}
\affil[2]{University of Oslo, Department of Geosciences, Oslo, 0316, Norway}

\bigskip
\goodbreak

\affil[*]{Corresponding author: Gebregergs, Hagos Gebrehiwet. E-mail: hagos.gebrehiwet@mu.edu.et}

\begin{abstract}
The frequency content of seismic data is changing with propagation depth due to intrinsic absorption.
This implies that the higher frequencies are highly attenuated, thus leading to a loss in resolution
of the seismic image. In addition, absorption anomalies, for example, caused by gas sands, will further
dim the seismic reconstruction. It is possible to correct for such absorption effects by employing so-called inverse Q-filtering (IQF). This is a filtering technique that tries to restore the loss of the higher frequencies due to propagation. Newer developments within IQF can be regarded as a migration type of algorithm, and such classes of techniques are studied in this paper. As seismic waves travel through the earth, the visco-elasticity of the earth's medium will cause energy dissipation and waveform distortion. This phenomenon is referred to as seismic absorption. In explaining the propagation of seismic wave in a given medium we explore the relationship between the pressure and displacement stresses. Therefore, by introducing an absorption function into the stress and strain relationship we derived a non-linear wave equation. We, then, employed a layered earth model to solve the non-linear wave equation. 
\end{abstract}

\begin{document}

\flushbottom
\maketitle

\thispagestyle{empty}

\section{Introduction}

Seismic method utilizes the propagation of waves through the earth. Propagation of seismic wave is affected
by the anelasticity and heterogeneity of the earth's subsurface. As a result, seismic waves get absorbed while propagating through the earth's subsurface, \cite{Futt1962, Kjart1979, Kolsky1956, Ricker1953, Strick1967}.
The absorption of seismic waves in the earth's subsurface causes both amplitude attenuation and velocity
dispersion of the reflections recorded on the surface leading to a loss of resolution thereby resulting in loss of information concerning targets of potential interest. We want therefore to try to remove the effect of absorption. In seismic data processing, absorption can be compensated for to enhance seismic
data resolution. In hydrocarbon reservoir description, seismic absorption
can be used as an important attribute to interpret for fluid units.

To remove the effect of absorption, the most common ones are time-variant
spectral whitening, time-variant deconvolution and inverse Q filtering. Time-variant spectral whitening
tries to recover the lost high-frequency energy by applying an exponential gain function to the seismic
data divided into frequency bands. Time-variant deconvolution implements absorption compensation by
means of a time-variant wavelet in a moving time window. Inverse Q filtering is a deterministic process
that requires knowledge of the quality factor Q. The chief difficulty to absorption compensation lies on the nonstationary characteristics generated by the absorption process in the seismic traces recorded. In an
attenuating medium, the shape of seismic pulse wavelet recorded in the beginning of the trace will
differ from the wavelet's shape recorded in the later times of the signal.

\cite{LJGelius1987} showed that application of inverse Q filtering on field data from a shallow gas reservoir in the North Sea generally improved the resolution, and improved the match between the synthetic and the field data on both sides of the well.  In order to recover a weak reflection of a target gas-layer underneath a strong coal-seam reflections, \cite{wang2004seismic} were able to show improvement of vertical resolution in migrated seismic sections applying migration with inverse Q filtering algorithm in a real data. In \cite{wang2009seismic} it is shown that application of inverse Q filter on real seismic data resulted in better matching of seismic impedance sections and well-log information, enhanced seismic resolution of a potential target reservoir, and better reservoir characterization like lateral heterogeneity of a carbonate gas reservoir. In \cite{oliveira2013l1} it is shown that inversion was capable of raising the resolution of real data that was affected by shallow gas accumulations without boosting the noise present in the data.

Inverse Q filtering algorithms can be designed based on downward wave propagation migration type approach \cite{Harg1991, Wang2006, wang2014absorption, zhao2018inverse, rao2019stabilised}. There are stability issues with these inverse Q filtering algorithms as we move to the high frequency bands. Usually, the dispersion and amplitude corrections are dealt separately. This is because these inverse Q filters are easily separated into both phase-only and amplitude-only components. Moreover, dispersion corrections of seismic data are related to phase-only changes and are always stable. On the other hand, attenuation corrections correspond to amplitude-only changes and raise stability issue, \cite{van2012bandwidth, zhao2018inverse, rao2019stabilised}.

A slightly different approach of "Inversion of Reflection Data" is presented by \cite{Nilsen1978}. Nilsen and Gjevik solved the non-linear wave equation for a pressure wave by introducing an absorption function $\Theta=\Theta(\alpha,\rho,\eta,k,\omega)$, a complex function of absorption coefficient $\alpha$, density $\rho$, viscosity $\eta$, wave number $k$ and frequency $\omega$. Similarly, in this paper, a layered earth model is implemented to solve the non-linear wave equation by introducing absorption function $Y(\omega,\tau)$, where two way traveltime $\tau$ and frequency $\omega$.

\section{Non-Linear Modelling with Absorption}

Seismic wave traveling in the earth experience amplitude attenuation and velocity dispersion due to the absorptive
property of the rocks of the subsurface. This absorptive behavior of earth can be modelled where each layer of the
earth is described by layer thickness $z$, velocity $v$, density $\rho$ and a quality factor $Q$. When using Q to describe seismic attenuation, it is often assumed that Q does not change with frequency in the seismic
frequency range (10 - 200Hz).

Let us assume a plane-wave (pressure stress, $P$) propagating along a vertical axis in a medium whose density $\rho$ and seismic velocity $v$ change eventually with depth. In a frequency domain, Newton's second law gives
\begin{equation}\label{eqc2.1}
    \frac{\partial P}{\partial z} = -\rho\omega^{2}W,
\end{equation}
where $W$ denote the displacement. \cite{Gjevik1976} assumed the functional relation between the stress and the strain fields to be linear and isotropic as
\begin{equation}\label{eqc2.2}
    P =  \rho v^{2}\frac{\partial W}{\partial z}.
\end{equation}
The wave speed of the propagating pressure stress depends on the medium elastic parameters $\lambda$ and $\mu$ as in
\begin{equation}\label{eqc2.2v}
    v =  \sqrt{\frac{\lambda + 2\mu}{\rho}}.
\end{equation}
However, equation~(\ref{eqc2.2}) does not include the absorption aspect of the seismic wave propagation. Therefore,
\cite{Nilsen1978} corrected for the absorption effect by assuming Kelvin Voigt model, \cite{Jaeger1962}, stress-strain relationship
\begin{equation}\label{eqc2.3}
    P =  \rho v^{2}\Theta^{2}\frac{\partial W}{\partial z},
\end{equation}
with $\Theta=\Theta(\alpha,\rho,\eta,k,\omega)$ is being a complex-valued function of absorption coefficient $\alpha$,
density $\rho$, viscosity $\eta$, wave number $k$ and frequency $\omega$.

In this paper, an absorption function $Y(\omega,\tau)$ has been introduced in the stress and strain relationship as
\begin{equation}\label{eqc2.4}
    P =  \rho v_{r}^{2}Y\frac{\partial W}{\partial z},
\end{equation}
where $v_{r}$ is the reference velocity which could be taken as group velocity in the case of dispersion. The absorption function  $Y(\omega,\tau)$ depends on depth (measured in two way traveltime $\tau$) and frequency $\omega$, with $Y=1$ being absorption free as in equation~(\ref{eqc2.2}), \cite{Gjevik1976}. Therefore, combining equations~(\ref{eqc2.1}) and (\ref{eqc2.4}) give Helmholtz equation, assuming constant density which is in agreement with the layered approach in this paper:
\begin{equation}\label{eqc2.5}
    \frac{\partial^{2}P}{\partial z^{2}} + \tilde{k}^{2} P = 0, \qquad \tilde{k}=\frac{\omega}{v_{r}\sqrt{Y}},
\end{equation}
with $\tilde{k}$ being a complex-valued wavenumber.

\section{Absorption Function and $Q$ Models}
Following \cite{Horton1959}, we introduced the notation
\begin{equation}\label{eqc2.6}
    Y(\omega,\tau) =  A(\omega,\tau)+iB(\omega,\tau),
\end{equation}
which depends on the different $Q$ models. Since the complex-valued wavenumber $\tilde{k}$ is in focus,
the following expression is now elaborated on by first order Taylor expansion:
\begin{eqnarray}\label{eqc2.7}
\nonumber \frac{1}{\sqrt{Y}} &=& \frac{1}{\sqrt{A + iB}}= A^{-1/2}\bigg[1 + i\frac{B}{A}\bigg]^{-1/2}
\cong A^{-1/2}\bigg[1 - i\frac{B}{2A}\bigg] \quad \mathrm{for} \quad A\gg B,\\
    \tilde{k} &=& \frac{\omega}{v_{r}\sqrt{Y}} = \frac{\omega}{v_{r}\sqrt{A + iB}}\cong
    \frac{\omega}{v_{r}}\bigg[\frac{1}{\sqrt{A}} -\frac{i}{2}\frac{B}{A\sqrt{A}}\bigg].
\end{eqnarray}

In the literature the complex-valued wavenumber $\tilde{k}$ is often written on the following form in case of absorption:
\begin{eqnarray}\label{eqc2.17}
 \nonumber \tilde{k} &=& \frac{\omega}{v(\omega)}\bigg[1 -\frac{i}{2Q}\bigg] = \frac{\omega}{v_{r}}+
    \bigg[\frac{\omega}{v(\omega)} - \frac{\omega}{v_{r}}\bigg]-i\alpha(\omega), \\
            &=& \frac{\omega}{v_{r}}+\varphi(\omega)-i\alpha(\omega), \qquad \alpha(\omega)=\frac{\omega}{2Qv(\omega)},
\end{eqnarray}
where $\alpha$ is absorption coefficient and $\varphi$ is phase of the 'absorption filter'.
In order to ensure causality, this filter should be minimum phase. For such a filter this relationship holds
\begin{equation}\label{eqc2.18}
    \varphi(\omega) = \aleph\big[\alpha(\omega)\big],
\end{equation}
with $\aleph$ denoting the Hilbert Transform. If we omit dispersion (put $\varphi=0$), the filter will
be non-causal. Equating equations~(\ref{eqc2.7}) and~(\ref{eqc2.17}) gives the relationships
\begin{equation}\label{eqc2.19}
    A=\bigg[\frac{v(\omega)}{v_{r}}\bigg]^{2}, \qquad B=\bigg[\frac{v(\omega)}{v_{r}}\bigg]^{2}\frac{1}{Q}.
\end{equation}

According \cite{Aki2002} to honor causality the relation
\begin{equation}\label{eqc2.20}
    \frac{\omega}{v(\omega)} - \frac{\omega}{v_{\infty}} = \aleph\bigg[\frac{\omega}{2Qv_{\infty}}\bigg],
\end{equation}
should be held while $v_{\infty}$ is the limit of the velocity function as $\omega$ approaches infinity.
Equation~(\ref{eqc2.20}) can be further approximated as
\begin{equation}\label{eqc2.21}
    \frac{\omega}{v(\omega)} - \frac{\omega}{v_{h}} = \aleph\bigg[\frac{\omega}{2Qv_{h}}\bigg],
\end{equation}
where $v_{h}$ is the velocity at the tuning frequency (125\,Hz in this paper) of the seismic band,
\cite{Wang2004}. Therefore, the complex-valued wavenumber
is accordingly adjusted as (compare with equation~(\ref{eqc2.17}))
\begin{equation}\label{eqc2.22}
    \tilde{k} = \frac{\omega}{v_{h}}+ \bigg[\frac{\omega}{v(\omega)} - \frac{\omega}{v_{h}}\bigg]-i\frac{\omega}{2Qv(\omega)}=
    \frac{\omega}{v_{h}}\bigg\{1+ \bigg[\frac{v_{h}}{v(\omega)} - 1\bigg]-i\frac{v_{h}}{2Qv(\omega)}\bigg\}
\end{equation}
and combined with a \cite{Kolsky1956} type of phase-velocity model of frequency independent Q ($Q\gg1$)
\begin{equation}\label{eqc2.23}
    v(\omega) = v_{h}\bigg(\frac{\omega}{\omega_{h}}\bigg)^{\gamma}, \qquad \gamma=(\pi Q)^{-1}
\end{equation}
gives the complex-valued wavenumber model
\begin{equation}\label{eqc2.24}
    \tilde{k} = \frac{\omega}{v_{h}}\bigg\{1+ \bigg[\bigg(\frac{\omega}{\omega_{h}}\bigg)^{-\gamma} - 1\bigg]-
    i\frac{1}{2Q}\bigg(\frac{\omega}{\omega_{h}}\bigg)^{\gamma}\bigg\},
\end{equation}
which has been employed by \cite{Wang2004}. From equations~(\ref{eqc2.19}) and~(\ref{eqc2.23}) it also follows that ($v_{r}=v_{h}$)
\begin{equation}\label{eqc2.25}
    A_{Wang}=\bigg[\frac{\omega}{\omega_{h}}\bigg]^{2\gamma}, \qquad B_{Wang}=\bigg[\frac{\omega}{\omega_{h}}\bigg]^{2\gamma}\frac{1}{Q}.
\end{equation}

Based on equation~(\ref{eqc2.20}), \cite{Kjart1979} proposed an alternative complex-valued wavenumber model
\begin{equation}\label{eqc2.26}
\tilde{k} = \frac{\omega}{v_{r}}+ \bigg[\frac{\omega}{v(\omega)} - \frac{\omega}{v_{r}}\bigg]-i\frac{\omega}{2Qv(\omega)}
\approx\frac{\omega}{v_{r}}+\aleph\bigg[\frac{\omega}{2Qv_{r}}\bigg]-i\frac{\omega}{2Qv_{r}}.
\end{equation}
Based on equations~(\ref{eqc2.19}) and~(\ref{eqc2.26}) it follows that with $v_{r}=v_{\infty}$
\begin{equation}\label{eqc2.27}
    A_{Kjar}=\bigg[1+\frac{1}{\omega}\aleph\bigg(\frac{\omega}{2Q}\bigg)\bigg]^{2},
    \qquad B_{Kjar}=A_{Kjar}^{3/2}\frac{1}{Q}.
\end{equation}
Finally, the dispersion-free and non-causal absorption Q model of \cite{Futt1962},
i.e. $v(\omega)=v_{r}$, corresponds to
\begin{equation}\label{eqc2.28}
    A_{no-disp}=1, \qquad B_{no-disp}=\frac{1}{Q}.
\end{equation}
If the quality factor is extremely large, $Q\rightarrow\infty$, then $(1/Q)\rightarrow0$. Hence, all the
mentioned Q-models become absorption free, that is $Y(\omega,\tau)\simeq1$.

\section{Plane Wave Solutions}
In case of a layer/medium with constant velocity and absorption function $Y$, plane-wave
solutions of equation~(\ref{eqc2.5}) can be written formally on the simple form (positive
z-axis pointing downwards)
\begin{equation}\label{eqc2.29}
    U(\omega,z)=a\exp[i\tilde{k}z], \qquad D(\omega,z)=b\exp[-i\tilde{k}z],
    \qquad \tilde{k}\equiv k-i\alpha \,(k,\, \alpha\geq0)
\end{equation}
with $U(\omega,z)$ and $D(\omega,z)$ representing respectively upward and downward propagating
components. Thus, the total field can be written
\begin{equation}\label{eqc2.30}
    P=U + D.
\end{equation}
From equation~(\ref{eqc2.1}) and by using equations~(\ref{eqc2.29}) and~(\ref{eqc2.30}) as well as
the expression of $\tilde{k}$ from equation~(\ref{eqc2.5}) gives it follows that
\begin{eqnarray}\label{eqc2.31}
\nonumber  \frac{\partial P}{\partial z}&=&\frac{\partial U}{\partial z}+\frac{\partial D}{\partial z}
=i\tilde{k}U-i\tilde{k}D=-i\tilde{k}[D-U],\\
W&=&-\frac{1}{\rho\omega^{2}}\frac{\partial P}{\partial z}=\frac{i}{\omega\rho v_{r}\sqrt{Y}}[D - U].
\end{eqnarray}

A depth-varying model can be assumed as the limit of an infinite number of infinitesimal layers. For
such a model where the acoustic impedance and the absorption function weakly inhomogeneous compared
to the wavelength of the wave, the relation in equation~(\ref{eqc2.31}) is also assumed to be valid,
\cite{Nilsen1978}. While making use of equation~(\ref{eqc2.4}), differentiating equation~(\ref{eqc2.31})
becomes
\begin{eqnarray}\label{eqc2.32}
\nonumber \frac{\partial W}{\partial z}&=&\frac{1}{\rho v_{r}^{2}Y}P=\frac{1}{\rho v_{r}^{2}Y}[U + D],\\
&=&\frac{i}{\omega\rho v_{r}\sqrt{Y}}\bigg[\frac{\partial D}{\partial z}-\frac{\partial U}{\partial z}\bigg]
+\frac{i}{\omega}[D-U]\frac{\partial}{\partial z}\bigg[\frac{1}{\rho v_{r}\sqrt{Y}}\bigg].
\end{eqnarray}

Further simplification of equation~(\ref{eqc2.32}) can be obtained by considering small to moderate absorption:
\begin{equation*}
\frac{\partial}{\partial z}\bigg[\frac{1}{\rho v_{r}\sqrt{Y}}\bigg]=-\frac{1}{\rho v_{r}\sqrt{Y}}
\bigg[\frac{1}{\sqrt{Y}}\frac{\partial\sqrt{Y}}{\partial z}+\frac{1}{\rho v_{r}}\frac{\partial(\rho v_{r})}{\partial z}\bigg].
\end{equation*}
For small to moderate absorption, $\frac{1}{\rho v_{r}}\frac{\partial[\rho v_{r}]}{\partial z}
\gg\frac{1}{\sqrt{Y}}\frac{\partial\sqrt{Y}}{\partial z}$:
\begin{eqnarray}
\frac{\partial}{\partial z}\bigg[\frac{1}{\rho v_{r}\sqrt{Y}}\bigg]&\cong&-\frac{1}{\rho v_{r}\sqrt{Y}}
\bigg[\frac{1}{\rho v_{r}}\frac{\partial(\rho v_{r})}{\partial z}\bigg]=-\frac{2}{\rho v_{r}\sqrt{Y}}\Upsilon(z), \label{eqc2.33a} \\
\Upsilon(z)&=&\frac{1}{2\rho v_{r}}\frac{\partial[\rho v_{r}]}{\partial z},\label{eqc2.33}
\end{eqnarray}
where $\Upsilon(z)$ represents the depth-dependent 'reflectivity' (reflectivity per depth unit). The approximation
$\frac{1}{\rho v_{r}}\frac{\partial[\rho v_{r}]}{\partial z}\gg\frac{1}{\sqrt{Y}}\frac{\partial\sqrt{Y}}{\partial z}$
holds for low frequency waves in reflection studies of rocks and sediments, \cite{Nilsen1978}. Therefore, using
equation~(\ref{eqc2.33a}), equation~(\ref{eqc2.32}) can be rearranged to give
\begin{equation}\label{eqc2.34a}
\frac{\partial D}{\partial z}-\frac{\partial U}{\partial z}=-\frac{i\omega}{v_{r}\sqrt{Y}}[D+U]+2\Upsilon(z)[D-U].
\end{equation}
Similarly, combination of equations~(\ref{eqc2.1}) and~(\ref{eqc2.30}) gives
\begin{equation}\label{eqc2.34}
\frac{\partial P}{\partial z}=-\rho\omega^{2}W \quad\Longrightarrow\quad
\frac{\partial D}{\partial z}+\frac{\partial U}{\partial z}=-\frac{i\omega}{v_{r}\sqrt{Y}}[D-U].
\end{equation}
And a main result is obtained now by combining equations~(\ref{eqc2.34a}) and~(\ref{eqc2.34}),
\begin{eqnarray}\label{eqc2.35}
\nonumber  \frac{\partial D}{\partial z} &=& -\frac{i\omega}{v_{r}\sqrt{Y}}D + \Upsilon(z)[D-U], \\
  \frac{\partial U}{\partial z} &=& \frac{i\omega}{v_{r}\sqrt{Y}}U - \Upsilon(z)[D-U].
\end{eqnarray}

To solve the above coupled equations~(\ref{eqc2.35}), let us now introduce the earth's response
described by the ratio $K=U/D$, \cite{Yilmaz2001}, and differentiate it with respect to depth
\begin{equation}\label{eqc2.36}
\frac{\partial K}{\partial z}=\frac{\partial [U/D]}{\partial z}
=\frac{1}{D}\frac{\partial U}{\partial z}-\frac{U}{D^{2}}\frac{\partial D}{\partial z}.
\end{equation}
Finally, by combining equations~(\ref{eqc2.35})~-~(\ref{eqc2.36}) gives the Ricatti equation,
\cite{Nilsen1978},
\begin{equation}\label{eqc2.37}
\frac{\partial K}{\partial z}=\frac{2i\omega}{v_{r}\sqrt{Y}}K-\Upsilon[1-K^{2}].
\end{equation}

Since vertically traveling waves are considered, the transformation from depth to two-way traveltime
is straightforward
\begin{eqnarray}\label{eqc2.38}
\tau&=&2\int_{0}^{z}\frac{dz}{v_{r}}, \qquad\Rightarrow\, d\tau=\frac{2}{v_{r}}dz \quad \mathrm{and}
\quad \frac{\partial}{\partial z}=\frac{2}{v_{r}}\frac{\partial}{\partial\tau}, \\
\nonumber \Upsilon(z) &=&\frac{1}{2\rho v_{r}}\frac{\partial[\rho v_{r}]}{\partial z}\equiv\frac{1}{2\rho v_{r}}
\bigg[\frac{2}{v_{r}}\frac{\partial[\rho v_{r}]}{\partial\tau}\bigg]=\frac{2}{v_{r}}
\bigg[\frac{1}{2\rho v_{r}}\frac{\partial[\rho v_{r}]}{\partial\tau}\bigg]=\frac{2}{v_{r}}r(\tau),\\
& & \Longrightarrow \quad r(\tau) = \frac{1}{2\rho v_{r}}\frac{\partial[\rho v_{r}]}{\partial\tau},
\end{eqnarray}
which gives the traveltime-version of equation~(\ref{eqc2.37})
\begin{eqnarray}\label{eqc2.39}
\nonumber \frac{2}{v_{r}}\frac{\partial K(\omega,\tau)}{\partial\tau}&=&
\frac{2i\omega}{v_{r}\sqrt{Y(\omega,\tau)}}K(\omega,\tau)-\frac{2}{v_{r}}r(\tau)[1-K^{2}(\omega,\tau)],\\
\frac{\partial K(\omega,\tau)}{\partial\tau}&=&\frac{i\omega}{\sqrt{Y(\omega,\tau)}}K(\omega,\tau)-r(\tau)[1-K^{2}(\omega,\tau)].
\end{eqnarray}
Equation~(\ref{eqc2.39}) is a non-linear wave equation due to multiples included as a result of weak
acoustic impedance contrast of layers in the model (of course due to strong acoustic impedance as well).
Its solution is the starting point of the forward and inversion algorithms. It should be solved in order
to apply for a stratified earth with each layer defined by a two-way traveltime $\tau_{j}$, for
$j = 0, 1, ..., N-1$, with $\tau_{0}=0$ and $\tau_{N}$ being the maximum record. In case of the
forward modelling, the plane-wave upward continuation from the top of the $j$th layer to surface
can be implemented recursively.

\section{Forward $Q$ Modelling - Non-Linear Case}
With
\begin{eqnarray}
\Phi(\omega,\tau)&=&\int_{0}^{\tau}\frac{\omega}{\sqrt{Y(\omega,\tau')}}\partial\tau'
\cong\int_{0}^{\tau}\bigg[\frac{\omega}{\sqrt{A}}-\frac{i\omega B}{A\sqrt{A}}\bigg]\partial\tau',\label{eqc2.1valerie}\\
\frac{\partial}{\partial\tau}\bigg[K(\omega,\tau)\exp\big[-i\Phi(\omega,\tau)\big]\bigg]&=&\bigg[\frac{\partial K(\omega,\tau)}{\partial\tau}+\frac{-i\omega}{\sqrt{Y(\omega,\tau)}}K(\omega,\tau)\bigg]\exp\big[-i\Phi(\omega,\tau)\big],
\label{eqc2.2valerie}
\end{eqnarray}
the Riccati equation of equation~(\ref{eqc2.39}) can be rewritten on the following form:

\begin{equation}\label{eqc2.40}
\frac{\partial}{\partial\tau}\bigg[K(\omega,\tau)\exp\big[-i\Phi(\omega,\tau)\big]\bigg]
=-r(\tau)\big[1-K^{2}(\omega,\tau)\big]\exp\big[-i\Phi(\omega,\tau)\big].
\end{equation}

\begin{figure}[htb!]
\subfloat[]{\includegraphics[width = 3in]{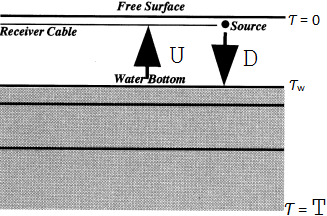}}
\hfill
\subfloat[]{\includegraphics[width = 3in]{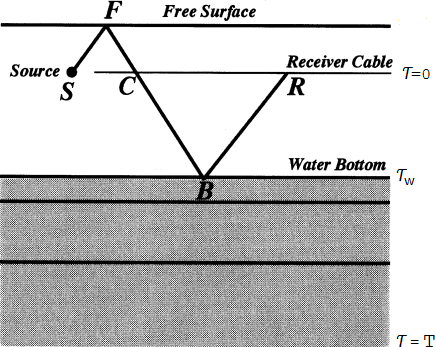}}
\caption{(a) 1D Earth model: D is downward while U is an upward propagating components. (b) Water-bottom multiples due to air-water reflection. (Adapted from \cite{Yilmaz2001}.)}
\label{fig1:2}
\end{figure}

Figure~\ref{fig1:2}a shows 1D stratified earth model whereby $T$ is the maximum thickness of the model in two-way
traveltime. As the source is at the top (water) layer, there are no up-going waves at the bottom layer. Therefore,
$U(\omega,T)=0$ when $\tau\geq T$ which implies $K(\omega,T)=0$. When trying to compute the earth's response
$K(\omega,\tau)$ at any depth $\tau$, we can apply the boundary condition $K(\omega,T)=0$ up on integration of
equation~(\ref{eqc2.40}):
\begin{equation}\label{eqc2.42}
    -K(\omega,\tau)\exp\big[-i\Phi(\omega,\tau)\big]=-\int_{\tau}^{T}r(\tau')\exp\big[-i\Phi(\omega,\tau')\big]\big[1-K^{2}(\omega,\tau')\big]\partial\tau'.
\end{equation}
Rearranging equation~(\ref{eqc2.42}) gives
\begin{equation}\label{eqc2.43}
    K(\omega,\tau)=\exp\big[i\Phi(\omega,\tau)\big]\int_{\tau}^{T}r(\tau')\exp\big[-i\Phi(\omega,\tau')\big]\big[1-K^{2}(\omega,\tau')\big]\partial\tau',
\end{equation}
where equation~(\ref{eqc2.43}) is now the starting point for a forward and inverse Q modelling algorithms.

With time step $\triangle\tau=4\times10^{-3}$ seconds, $j\times\triangle\tau$, the integral part of equation~(\ref{eqc2.43}) can be solved using finite difference method:
\begin{equation}\label{eqc2.44}
K(\omega,\tau)=\exp\big[i\Phi(\omega,\tau)\big]\bigg[\int_{\tau}^{\tau+\triangle\tau}+\int_{\tau+\triangle\tau}^{\tau+2\triangle\tau}+
...+\int_{(N-3)\triangle\tau}^{(N-2)\triangle\tau}+\int_{(N-2)\triangle\tau}^{(N-1)\triangle\tau}\bigg]\bigg[\qquad\bigg]\partial\tau',
\end{equation}
by making use of the trapezoid integral
$$\int_{\tau}^{\tau+\triangle\tau}r(\tau')\exp\big[-i\Phi(\omega,\tau')\big]\big[1-K^{2}(\omega,\tau')\big]\partial\tau'\cong$$
$$\frac{\triangle\tau}{2}\bigg[r(\tau+\triangle\tau)\exp\big[-i\Phi(\omega,\tau+\triangle\tau)\big]\big[1-K^{2}(\omega,\tau+\triangle\tau)\big]
+ r(\tau)\exp\big[-i\Phi(\omega,\tau)\big]\big[1-K^{2}(\omega,\tau)\big]\bigg].$$
Assuming a discretization in $\tau$ (sample interval $\triangle\tau$ and total of $N$ points), calculation of
seismogram $K(\omega,\tau)$ in upward direction is done starting at maximum time $T=(N-1)\triangle\tau$.
Introducing the following notation for convenience
\begin{eqnarray*}\label{eqc2.45}
   \tau_{j} &=& j\times\triangle\tau  \qquad \& \qquad j = N-1, \,N-2, \,..., \,0, \\
 \omega_{l} &=& 2\pi\times l\times\triangle f \qquad \& \qquad \triangle f=\frac{1}{N\times\triangle\tau},\\
K_{l,j}^{n} &=& K^{n}(\omega_{l},\tau_{j}), \qquad \chi_{l,j} = \exp\big[-i\Phi(\omega_{l},\tau_{j})\big],
 \qquad r_{j} = r(\tau_{j}),
\end{eqnarray*}
where the superscript $n$ implies iteration number. The iteration number $n$ represents the number of times a
computation cycle repeated, applying each time the previous result, to get successively closer approximations
to the solution of equation~(\ref{eqc2.44}). This is because equation~(\ref{eqc2.44}) is non-linear. Therefore,
applying the trapezoidal integral in equation~(\ref{eqc2.44}), the $(n+1)^{th}$ iteration is
\begin{eqnarray}\label{eqc2.46}
\nonumber K_{l,j=J}^{n+1} &=& \chi_{l,j=J}^{-1}\bigg\{\frac{\triangle\tau}{2}\bigg\}\bigg\{r_{J}\chi_{l,J}
[1-(K_{l,J}^{n})^{2}]+2\sum_{j=N-2}^{J+1}r_{j}\chi_{l,j}[1-(K_{l,j}^{n})^{2}]\\
&\qquad& \qquad\qquad\qquad\qquad + r_{N-1}\chi_{l,N-1}[1-(K_{l,N-1}^{n})^{2}]\bigg\},\\
\nonumber K_{l,N-1}^{n} &=& K^{n}(\omega_{l},T)=0, \quad \mathrm{boundary \, condition}.
\end{eqnarray}
The trapezoid approximation enables us to calculate the seismogram at each $J = 0,1, 2, ...., N-1$ corresponding to the two-way traveltime $\tau_{J} = J\times\triangle\tau$ with the seismogram corresponds to the solution $J = 0$, $\tau_{0} = 0\times\triangle\tau = 0$, is at the surface.

By dropping the $K^{2}$ in right-hand-side of equation~(\ref{eqc2.46}), a linear-forward wave
propagation model is obtained,
\begin{eqnarray}\label{eqch3.10}
\nonumber K_{l,j=J} &=& \chi_{l,j=J}^{-1}\bigg\{\frac{\triangle\tau}{2}\bigg\}\bigg\{r_{J}\chi_{l,J}+
2\sum_{j=N-2}^{J+1}r_{j}\chi_{l,j} + r_{N-1}\chi_{l,N-1}\bigg\}.\\
\end{eqnarray}
Equation~(\ref{eqch3.10}) is similar to the downward wave propagation model discussed in
\cite{montana2004compensating, Wang2006, van2012bandwidth}.

So far we have considered the source and receiver are at the surface as is shown in figure~\ref{fig1:2}a. This works
whether the first layer is water or not. Let us consider the source and receiver are deeper in the water layer. Now, we have to take into account surface-related multiples. Figure~\ref{fig1:2}b shows water-bottom multiples due to air-water reflection.

Assuming that $\tau_{w}$ represents two-way vertical travel time in the water layer, total field $P_{l}=P(\omega_{l})$
recorded at the receiver (including multiples) can then be written as ('$R$' being the reflection coefficient of the seafloor)
\begin{eqnarray}\label{eqc2.47}
\nonumber  P_{l}  &=& K_{l,j=0}\bigg[1-R\exp(-i\omega_{l}\tau_{w})+R\exp(-2i\omega_{l}\tau_{w})+....\bigg] \\
                  &=& \frac{K_{l,j=0}}{1+R\exp(-i\omega_{l}\tau_{w})}.
\end{eqnarray}
The final result in time, the modelled seismogram $k(t,0)$, is obtained after an inverse fourier transform ($FT^{-1}$). And for $J=0$, the matrix representation of equation~(\ref{eqc2.46}), $K^{2}$ being $(K^{n})^{2}$, is as follows:

\bigskip
\goodbreak
\begin{equation*}
\rotatebox{90}{$
%\begin{sideways}{$
\left[
  \begin{array}{c}
    K^{n+1}_{0,0} \\
    K^{n+1}_{1,0} \\
    . \\
    . \\
    . \\
    . \\
    K^{n+1}_{N-1,0} \\
  \end{array}
\right]=\frac{\triangle\tau}{2}
\left[
 \begin{array}{cccccc}
 \big[1-K^{2}_{0,0}\big]  & 2\chi_{0,1}\big[1-K^{2}_{0,2}\big] & . & 2\chi_{0,N-2}\big[1-K^{2}_{0,N-2}\big] & \chi_{0,N-1}\big[1-K^{2}_{0,N-1}\big] \\
 \big[1-K^{2}_{1,0}\big]  & 2\chi_{2,1}\big[1-K^{2}_{2,2}\big] & . & 2\chi_{1,N-2}\big[1-K^{2}_{1,N-2}\big] & \chi_{1,N-1}\big[1-K^{2}_{1,N-1}\big] \\
 . & .  &  .  & .  & .\\
 . & .  &  .  & .  & .\\
 . & .  &  .  & .  & .\\
 . & .  &  .  & .  & .\\
 \big[1-K^{2}_{N-1,0}\big]  & 2\chi_{N,1}\big[1-K^{2}_{N,2}\big] & . & 2\chi_{N-1,N-2}\big[1-K^{2}_{N-1,N-2}\big] & \chi_{N-1,N-1}\big[1-K^{2}_{N-1,N-1}\big] \\
 \end{array}
 \right]
 \left[
\begin{array}{c}
r_{0}\\
r_{1} \\
. \\
. \\
. \\
r_{N-2} \\
r_{N-1} \\
\end{array}
\right].
$}
%\end{sideways}
\end{equation*}

\section{Inverse $Q$ Filtering - Non-Linear Case}
In the limit $\tau\rightarrow0$  meaning that the receivers are at the surface, equation~(\ref{eqc2.43}) gives
the 'seismogram in frequency domain':
\begin{equation}\label{eqc2.48}
K(\omega,0) = \int_{0}^{T}r(\tau')\exp\big[-i\Phi(\omega,\tau')\big]\big[1-K^{2}(\omega,\tau')\big]\partial\tau'.
\end{equation}
Introducing 'reflectivity' series, specifically 'reflectivity per depth unit series' in this paper,
\begin{equation}\label{eqc2.49}
r(\tau') = \triangle\tau\sum_{j=0}^{N-1}r_{j}\delta(\tau'-j\times\triangle\tau) \qquad \mathrm{with} \qquad T=(N-1)\times\triangle\tau,
\end{equation}
equation~(\ref{eqc2.48}) can be written as
\begin{equation}\label{eqc2.50}
K(\omega,0) = \sum_{j=0}^{N-1}r_{j}\exp\big[-i\Phi(\omega,j\times\triangle\tau)\big]
\big[1-K^{2}(\omega,j\times\triangle\tau)\big]\triangle\tau.
\end{equation}
If we drop $K^{2}(\omega,j\times\triangle\tau)$ of the right hand side of equation~(\ref{eqc2.50}) that is multiples
are not included, an expression similar to that of \cite{oliveira2013l1} achieved, which in our case is linear.

Originally, seismogram is recorded in time-domain and sampled with a total of $NT$ samples. Fourier transform of
the data will give the same number of monochromatic seismograms and equation~(\ref{eqc2.50}) can be rewritten in
a matrix system:

\begin{equation}\label{eqc2.51}
\left[
  \begin{array}{c}
    K^{n+1}(0,0) \\
    K^{n+1}(1,0) \\
    . \\
    . \\
    . \\
    K^{n+1}(N-1,0) \\
  \end{array}
\right]=
\end{equation}
\begin{equation*}
 \triangle\tau\left[
 \begin{array}{cccccc}
 \chi(0,0)\big[1-(K^{n})^{2}\big]  & ... & \chi(0,N-1)\big[1-(K^{n})^{2}\big] \\
 \chi(1,0)\big[1-(K^{n})^{2}\big]  & ... & \chi(1,N-1)\big[1-(K^{n})^{2}\big] \\
 . & ... & . \\
 . & ... & . \\
 . & ... & . \\
 \chi(N-1,0)\big[1-(K^{n})^{2}\big] & ... & \chi(N-1,N-1)\big[1-(K^{n})^{2}\big] \\
 \end{array}
 \right]
\left[
\begin{array}{c}
r_{0}^{n} \\
r_{1}^{n} \\
. \\
. \\
. \\
r_{N-1}^{n} \\
\end{array}
\right],
\end{equation*}
where
\begin{equation*}
    \chi_{l,j} = \exp\big[-i\Phi(\omega_{l},\tau_{j})\big], \qquad \omega_{l}=2\pi\times l\times\triangle f
    \quad \mathrm{and} \quad \tau_{j} = j\times\triangle\tau.
\end{equation*}

For a given iteration number $n$, we can compute for the corresponding reflectivity series by solving equation~(\ref{eqc2.51}) employing standard least-squares inversion. Since iteration is repeating a computation cycle applying each time the previous result to get successively closer approximations to the solution, we start by taking $(K^{0}(\omega))^{2}=0$ for $n=0$. After a new estimate of the reflectivity series has been obtained, an update of $(K_{j}^{n}(\omega))^{2}$ can be obtained by solving the forward problem in equation~(\ref{eqc2.46}). Iterations are carried out until the relative change in reflectivity is below a certain user set threshold.

To apply least-squares, equation~(\ref{eqc2.51}) can be written in vector and matrix notation in short as
\begin{equation}\label{eqc2.52}
    \vec{K}=M\vec{r},
\end{equation}
where
\begin{equation*}
\vec{K}=\left[
  \begin{array}{c}
    K^{n+1}(0,0) \\
    K^{n+1}(1,0) \\
    . \\
    . \\
    . \\
    K^{n+1}(N-1,0) \\
  \end{array}
\right],\quad\quad\quad\quad
\vec{r}=\left[
\begin{array}{c}
r_{0}^{n} \\
r_{1}^{n} \\
. \\
. \\
. \\
r_{N-1}^{n} \\
\end{array}
\right]\quad \mathrm{and}
\end{equation*}

$$
M=\triangle\tau\left[
 \begin{array}{cccccc}
 \chi(0,0)\big[1-(K^{n})^{2}\big]  & ... & \chi(0,N-1)\big[1-(K^{n})^{2}\big] \\
 \chi(1,0)\big[1-(K^{n})^{2}\big]  & ... & \chi(1,N-1)\big[1-(K^{n})^{2}\big] \\
 . & ... & . \\
 . & ... & . \\
 . & ... & . \\
 \chi(N-1,0)\big[1-(K^{n})^{2}\big] & ... & \chi(N-1,N-1)\big[1-(K^{n})^{2}\big] \\
 \end{array}
 \right].$$

In general, there are more observations than model parameters. Let $\vec{K}$ be the desired
seismic output data while the actual output from equation~(\ref{eqc2.51}) is $\vec{S}=M\vec{r}$.
And the cumulative squared difference,
\begin{equation}\label{eqc2.54}
\nonumber \Sigma=(\vec{K}-M\vec{r})^{T\ast}(\vec{K}-M\vec{r})= \vec{K}^{T\ast}\vec{K}-\vec{K}^{T\ast}M\vec{r}-\vec{r}^{T\ast}M^{T\ast}\vec{K}+\vec{r}^{T\ast}M^{T\ast}M\vec{r}.% \\
\end{equation}
We want to estimate a reflectivity per depth unit series $\vec{r}$ given that
\begin{equation}\label{eqc2.54b}
    \frac{\partial\Sigma}{\partial\vec{r}}=0 \Longrightarrow -\vec{K}^{T\ast}M+\vec{r}^{T\ast}M^{T\ast}M=0.
\end{equation}
Because $\vec{r}^{T\ast}$ is complex valued, $\partial\vec{r}^{T\ast}/\partial\vec{r}=0$. Thus, applying matrix transpose and rearranging the terms of equation~(\ref{eqc2.54b})
\begin{equation}\label{eqc2.55}
\vec{r}=(M^{T\ast}M)^{-1}M^{T\ast}\vec{K}.
\end{equation}
Subsequently, it is also possible to compute the acoustic impedance $AI(\tau)=\rho(\tau)v_{r}(\tau)$. Following
equation~(\ref{eqc2.38})
\begin{equation*}\label{eqc2.56}
 \quad r(\tau) = \frac{1}{2\rho v_{r}}\frac{\partial(\rho v_{r})}{\partial\tau},  \Longrightarrow
 \rho(\tau)v_{r}(\tau) = \rho(0)v_{r}(0)\exp\bigg[2\int_{0}^{\tau}r(\tau')d\tau'\bigg],
\end{equation*}
and applying equation~(\ref{eqc2.43}), the acoustic impedance $AI_{j}=\rho(\tau_{j})v_{r}(\tau_{j})$ at $\tau_{j}$ from the surface can be put as
\begin{equation}\label{eqc2.56}
 AI_{j} = AI_{0}\exp\bigg[2\triangle\tau\sum_{i=0}^{j}r_{j}\bigg]\quad \mathrm{for}\quad j=0,1,2,..., N-1.
\end{equation}
$AI_{0}$ is easily known since the first layer is sea water in marine seismic.

Finally we can compute the reflectivity series $R_{j}$ as in
\begin{equation}\label{eqc2.57}
 R_{j} = \frac{AI_{j+1} - AI_{j}}{AI_{j+1} + AI_{j}}=
 \frac{\exp\big[2\triangle\tau r_{j+1}\big]-1}{\exp\big[2\triangle\tau r_{j+1}\big]+1}\quad \mathrm{for}\quad j=0,1,2,..., N-1.
\end{equation}
This will enable us to compare and contrast our results with other inversion methods as many of those inversion methods compute reflectivity series. Indeed, both the reflectivity series $R_{j}$ and reflectivity per unit depth $r_{j}$ (in this paper work) show acoustic impedance contrast. But they differ both in magnitude and dimensionality.

\section{Conclusion}

In this paper our main aim is to deal with compensation of absorption effects in seismic data by employing so-called inverse Q-filtering (IQF). This is a filtering technique that tries to restore the loss of higher frequencies due to propagation. It helps to improve the resolution of seismic image, and to balance the
frequency contents.

In layered approach for Q compensation, generally, we only require absorptive information for major layers.
Formulating absorption compensation as an inverse problem, and solving the inverse problem iteratively, the resolution of seismic data can be upgraded step by step while overcoming the instability problem which is a natural drawback of common inverse Q filtering. The inversion procedure is formulated on the basis that the seismogram at the surface $K(\omega,0)$ is a known quantity, while the reflectivity per depth unit series $r(\tau)$ is the unknown which is
to be determined. This seismogram $K(\omega,0)$ can be field or synthetic seismic data.

\newpage
\bibliography{main-blx}

\end{document}